\documentclass[pra,twocolumn,floatfix,nofootinbib]{revtex4-1}
\usepackage{float,epsfig}
\usepackage{color}
\usepackage{graphicx}
\usepackage{mathtools}
\RequirePackage{epstopdf,adjustbox}
\usepackage{amsthm, amsmath, amssymb, braket, tikz, algorithm, algorithmic}
\graphicspath{{Images/}}
\usepackage[colorlinks = True]{hyperref}
\usepackage{ulem}
\usepackage{soul}
\usepackage[toc]{appendix}
\begin{document}

\newtheorem{theorem}{\bf Theorem}[section]
\newtheorem{proposition}[theorem]{\bf Proposition}
\newtheorem{definition}[theorem]{\bf Definition}
\newtheorem{corollary}[theorem]{\bf Corollary}
\newtheorem{remark}[theorem]{\bf Remark}
\newtheorem{lemma}[theorem]{\bf Lemma}
\newcommand{\nrm}[1]{|\!|\!| {#1} |\!|\!|}

\newcommand{\ba}{\begin{array}}
\newcommand{\ea}{\end{array}}
\newcommand{\von}{\vskip 1ex}
\newcommand{\vone}{\vskip 2ex}
\newcommand{\vtwo}{\vskip 4ex}
\newcommand{\dm}[1]{ {\displaystyle{#1} } }

\newcommand{\be}{\begin{equation}}
\newcommand{\ee}{\end{equation}}
\newcommand{\beano}{\begin{eqnarray*}}
\newcommand{\eeano}{\end{eqnarray*}}
\newcommand{\inp}[2]{\langle {#1} ,\,{#2} \rangle}
\def\bmatrix#1{\left[ \begin{matrix} #1 \end{matrix} \right]}
\def \noin{\noindent}
\newcommand{\evenindex}{\Pi_e}
\newcommand{\tb}[1]{\textcolor{magenta}{ #1}}
\newcommand{\noteba}[1]{\textcolor{magenta}{\bf\small BA: #1}}
\newcommand{\tg}[1]{\textcolor{green}{ #1}}


\def \R{{\mathbb R}}
\def \C{{\mathbb C}}
\def \F{{\mathbb F}}
\def \J{{\mathbb J}}
\def \L{\mathcal{L}}
\def \H{{\mathcal H}}
\def \Q{{\mathcal Q}}
\def \Tr{\mathrm{Tr}}
\def \diag{\mathrm{diag}}
\def \bin{\mathrm{bin}}
\def \col{\mathrm{col}}

\title{Perfect state transfer on hypercubes and its implementation using superconducting qubits}
\author{Siddhant Singh$^1$}
\email{siddhant.singh@iitkgp.ac.in}
\author{Bibhas Adhikari$^{2,3}$}
\email{bibhas@maths.iitkgp.ac.in}
\author{Supriyo Dutta$^3$}
\email{dosupriyo@iitkgp.ac.in}
\author{David Zueco$^4$}
\email{dzueco@unizar.es}
\affiliation{$^1$Department of Physics, Indian Institute of Technology Kharagpur, Kharagpur 721302, India}
\affiliation{$^2$Department of Mathematics, Indian Institute of Technology Kharagpur, Kharagpur 721302, India}
\affiliation{$^3$Center for Theoretical Studies, Indian Institute of Technology Kharagpur, Kharagpur 721302, India}
\affiliation {$^4$Instituto de Nanociencia y Materiales de Aragón (INMA) and Departamento de Física de la Materia Condensada,
  CSIC-Universidad de Zaragoza, Zaragoza 50009,
  Spain}


\date{\today}

\begin{abstract}
We propose a protocol for perfect state transfer between any pair of vertices in a hypercube. Given a pair of distinct vertices in the hypercube we determine a sub-hypercube that contains the pair of vertices as antipodal vertices. Then a switching process is introduced for determining the sub-hypercube of a memory enhanced hypercube that facilitates perfect state transfer between the desired pair of vertices. Furthermore, we propose a physical architecture for the pretty good state transfer implementation of our switching protocol with fidelity arbitrary close to unity, using superconducting transmon qubits with tunable couplings. The switching is realised by the control over the effective coupling between the qubits resulting from the effect of ancilla qubit couplers for the graph edges. We also report an error bound on the fidelity of state transfer due to faulty implementation of our protocol.
\end{abstract}

\pacs{Valid PACS appear here}
\maketitle


\section{Introduction}
    In quantum computation, it is often required to transfer an arbitrary quantum state from one location to another \cite{ref:48}. 
    Especially in large scale QIP, this is an important task, connecting 
     two sites that may belong to the same or different quantum processors. 
    The latter is nontrivial for many quantum information processing (QIP) realizations, such as, solid state quantum computing and superconducting quantum computing \cite{ref:77,ref:75,ref:71,ref:72,ref:73,ref:74,ref:23}. 
    It is also very important to find the systems that support this quantum information exchange between distant sites to realize this phenomenon. For the short distance communication (such as adjacent quantum processors), methods for interfacing different kinds of physical systems are much required, for example, ion traps \cite{ref:39,ref:34}, superconducting circuits \cite{ref:41,ref:42}, boson lattices \cite{ref:76}, etc. The task of state transfer is incorporated with the idea of reducing the manipulation required to communicate between distant computational qubits in a large scale quantum computer \cite{ref:54,ref:49}. Scalability of quantum processors is a deep concern in the development of quantum computing hardware \cite{ref:26,ref:27,ref:28}. This is essential for determining how good is an architecture for quantum information processing \cite{ref:30}.

    
    Quantum state transfer with 100\% fidelity is known as perfect state transfer (PST) and this idea using interacting spin-1/2 particles was first proposed in \cite{ref:1}. This is established by utilizing a combinatorial graph structure as a platform for actual quantum network in the first excitation subspace of multi-qubit system \cite{ref:15,ref:2}. 
    In general, this involves mixed states of the network qubits \cite{ref:57}, however, showing PST for pure states in a graph suffices to prove the phenomenon. PST can be used in entanglement transfer, quantum communication, signal amplification, quantum information recovery and implementation of universal quantum computation \cite{ref:18,ref:49,ref:60,ref:61}. It is important to find classes of graphs where PST is possible and equally important to find graphs where it does \cite{ref:58,ref:12}, in order to classify how good is an architecture for quantum information processing. The idea of pretty good state transfer is also studied, where the transfer fidelity is less than unity but occurring on a large number of graphs that support state transfer \cite{ref:66,ref:67,ref:68,ref:79,ref:83} and in general these graphs can be weighted \cite{ref:65}. More general graphs such as signed graphs \cite{ref:50} and oriented graphs \cite{ref:19} are also studied. PST for qudits has also been classified for some networks \cite{ref:53,ref:21}. One of the big challenges for scalable quantum architecture is the imperfect two-qubit interaction. For PST with maximum fidelity, the pairwise interaction should be improved for large-scale quantum processors \cite{ref:31}. Different physical systems for quantum computation have different advantages, such as, high-fidelity and control in ion-traps \cite{ref:35,ref:34} versus the scalability of superconducting circuits \cite{ref:33,ref:23,ref:32}. PST was demonstrated in the latter, with tunable qubit couplings \cite{ref:78}. Here, we propose a protocol for large scale quantum processors with support on hypercube network.

    The scheme established in \cite{ref:4} and \cite{ref:5} allows PST over arbitrary long distances. 
    Here the shortcoming is that the PST is possible only between antipodal vertices. A switching technique is proposed in \cite{ref:56} where in a complete graph $K_n$, switching off one link establishes PST in non-adjacent qubits. This enables PST for more vertices but still does not enable routing to different vertices and there is no scalability, the graph remains fixed. One attempt at switching and routing is proposed in \cite{ref:59} which involves creating new edges and coupling for qubits, however, is still not scalable. Various other works describe methods of routing of excitations in spin chains \cite{ref:80,ref:81,ref:82}, limited to one dimension. In this work, we resolve this problem through our hypercube switching scheme. The goal of this paper is three fold: $1)$ showing perfect state transfer is possible between any pair of vertices in an $n$-dimensional hypercube by introducing a concept of switching on  and off of edges of the hypercube, $2)$ defining an effective Hamiltonian that can implement the sub-hypercube architecture with the proposed switching, $3)$ finding an error bound of the fidelity of the PST for an inaccurate implementation of the sub-hypercube in an experimental setup.

   The rest of the paper is organized as follows.
   We recall some preliminary results in section \ref{sec:preliminary} that are needed to establish our results. In section \ref{sec:hcube}, given a pair of distinct vertices in a $n$-dimensional  hypercube $\Q_n,$ a unique sub-hypercube $\Q_d, d\leq n$ is determined such that the given pair of vertices are antipodal vertices of $\Q_d.$ We also describe how a memory enhanced hypercube enables to identify the vertices of the sub-hypercube. Consequently, the perfect state transfer is established between those pair of vertices. A proposal for the implementation of our switching protocol is presented in section \ref{sec:implementation} using superconducting transmon qubits.
    
    

    \section{Preliminaries}
    \label{sec:preliminary}
	A graph is an ordered pair $G = (V, E)$, where $V$ denotes the set of vertices (or nodes), and $E \subseteq \{ (u, v) \in V\times V |  u \neq v\}$ denotes the set of edges, which are unordered pairs of vertices. Let $I\subseteq V.$ Then a  subgraph of the graph $G$ defined by the vertex set $I$ is called an \textit{induced subgraph} if two distinct vertices in $I$ are linked by an edge in the subgraph if and only if they are linked by an edge in $G$ \cite{west2001introduction}. Obviously, induced subgraph defined by a set of vertices is unique. 
	
	Let $G$ be a graph with $n$ vertices with $n=|V|$ and $V=\{0, 1, 2,\hdots, n-1\}.$ Then the adjacency matrix $A(G)=[a_{ij}]$ associated with $G$ is defined as  $a_{ij}=1$ if $(i, j)\in E$ and $0$ otherwise. Obviously, $A(G)$ is a symmetric matrix of order $n\times n.$ Then the degree of $i\in V$ is defined as $\mbox{deg}(i)=\sum_{j=0}^{n-1} a_{ij}.$ The degree matrix $D(G)=\mbox{diag}(d_0, \hdots, d_{n-1})$ of $G$ is a diagonal matrix where $d_i=\mbox{deg}(i), 0\leq i \leq n-1.$ The graph Laplacian matrix $L(G)$ associated with the  graph $G$ is defined as $L(G)=D(G)-A(G).$ These two matrices encode the structure of the graph, and determine the architecture by determining qubits' mutual couplings and connectivity.

	PST of a state in a many-qubit system is formulated by a combinatorial graph in which the edges of the graph represents coupling of qubits. This is interpreted by beginning with a single-qubit state (generally mixed) $\rho_{\text{qubit}}^u$ at some site $u$, and $\rho_{\text{in}}$ taken as the state of the rest of the system, and after evolution for some finite time $t_0$, with an interaction Hamiltonian $H$, the final evolved state
    \begin{equation}
    \label{eqn:pstgeneral}
    e^{-iHt_0/\hbar}(\rho_{\text{qubit}}^u\otimes \rho_{\text{in}})e^{iHt_0/\hbar}=\rho_{\text{qubit}}^v\otimes \rho_{\text{out}}
    \end{equation}
    is obtained, thereby transmitting the respective qubit state to another desired vertex $v$ of the graph. In general, $\rho_{\text{qubit}}$ is a density matrix, however, in this paper we consider that it corresponds to a pure state. The most simplified case for such realization is the one-dimensional chain of qubits.
    
   There are two kinds of well known interaction Hamiltonian for the pairwise interactions defined by the edges between the qubits which are placed at the vertices of a graph. The first is the XY model in two spatial degrees of freedom,
\begin{equation}
\label{eqn:XYmodel}
\begin{split}
        \frac{H_{XY}}{\hbar} & =\sum_{(i,j)\in E(G)}J_{ij}\left(\sigma^x_i\sigma^x_j+\sigma^y_i\sigma^y_j\right)\\
        &=\sum_{(i,j)\in E(G)}2J_{ij}\left(\sigma^+_i\sigma^-_j+\sigma^-_i\sigma^+_j\right),
\end{split}
\end{equation}
where $\sigma_i^\pm$ are the ladder operators acting on the qubit placed at vertex $i$ such that $\sigma^{\pm}_{i}=\sigma_i^x\pm i \sigma^y_i$, with $\sigma^{x,y}_i$ are the Pauli matrices. The second connection is via the three-dimensional Heisenberg model,
\begin{equation}
    \frac{H_{Hei}}{\hbar}=-\sum_{(i,j)\in E(G)}J_{ij}\Vec{\sigma}_i\cdot \Vec{\sigma}_j+\sum_jB_j\sigma^z_j
\end{equation}
where, $\Vec{\sigma}_i$ is Pauli matrix vector $\Vec{\sigma}_i=(\sigma^x_i,\sigma^y_i,\sigma^z_i)$ for the $i$th spin and $I_i$ is the identity operator for the $i$th vertex. 

In this paper, we consider the general many-body XY-coupling as well as the Heisenberg Hamiltonian and consider the coupling strength $J_{ij}$ to be a real parameter that can be continuously changed. In addition, we consider the local fields $B_j$ to tune the diagonal terms of the Hamiltonian such that it coincides with the Laplacian of the graph. This special choice is always possible \cite{ref:4}. We also emphasize that these two coupling Hamiltonian need not necessarily correspond to spin-1/2 particle interaction. In the first excitation subspace for the qubits, XY-coupling Hamiltonian and Heisenberg Hamiltonian action is equivalent to adjacency and Laplacian action respectively, for the corresponding graph, in the vertex space \cite{ref:15}. Initially, the system is in its ground state $|0\rangle = |000...0\rangle$, where the ket $|0\rangle$ denotes the single qubit ground state. We have the graph vertex-space states $|i\rangle = |00...010....0\rangle$ ($i = 1,2,..,n$), in which the qubit at the $i$th site is in the first excited state $|1\rangle$. To start the PST procedure, $A$ encodes an unknown (and arbitrary) state $|\psi_{in}\rangle = \cos(\theta/2)|0\rangle + e^{i\phi} \sin(\theta/2)|1\rangle$ at site $A$ in the graph and lets the system evolve freely for a finite time $t=t_0$ at which, the quantum state localizes at another site $B$ in the many-body system. This free quantum evolution of the entire network is precisely the quantum walk on the corresponding graph $G$ corresponding to the many-body network.

To define quantum walk and state transfer on a  graph $G=(V,E)$ of $n=|V|$ vertices, $n$-qubit states are considered that are localized at the vertices of the graph, equivalently the excitation space isomorphic to $\C^{|V|}.$ A continuous time quantum walk on a graph $G$ is the Schr\"{o}dinger evolution of the graph composite state with the graph adjacency matrix $A(G)$ as the Hamiltonian \cite{kendon2011perfect}. If $\ket{\zeta(0)} \in \mathbb{C}^{|V|}$ is the initial quantum state, then the evolution of the quantum walk is given by
\begin{equation}\label{adj:dyn}
			\ket{\zeta(t)} = \exp(-itA(G))\ket{\zeta(0)},
\end{equation} 
The probability for getting the quantum state localised at the vertex $v$ at time $t$ is given by $|\braket{v|\zeta(t)}|^2$. $G$ has a PST from vertex $u$ to vertex $v$ at finite time $t_0$ if
		\begin{equation}
			|\braket{v| \exp(-it_0A(G)) | u}| = 1.
		\end{equation}
		This is the same condition for perfect state transfer expressed in graph theoretic fashion \cite{ref:7} and implies Eqn. (\ref{eqn:pstgeneral}). The graph $G$ allows a perfect state transfer from the vertex $u$ to $v$ if the $(u,v)$-th term of $\exp(-itA(G))$ has magnitude $1$. Similarly, Laplacian PST is when the $(u,v)$-th term of $\exp(-itL(G))$ has magnitude $1$

 Some well-known examples of graphs which allow perfect state transfer over long distances are  \cite{ref:4,ref:5} :
		\begin{enumerate}
		\label{knownPST}
			\item 
				The complete graph $K_2$ with two vertices allow perfect state transfer between its vertices in time $t_0=\pi/2$ (in the units of energy inverse).
			\item
				The path graph $P_3$ has perfect state transfer between its end vertices  in time $t_0=\pi/\sqrt{2}$ (in the units of energy inverse).
			\item
				The hypercube of any order has perfect state transfer between its antipodal points in the same time $\pi/2$. Besides, on any order of Cartesian product of $P_3$ has PST between its antipodal vertices in the same time $\pi/\sqrt{2}$.
		\end{enumerate}	
Above three results hold both for the XY-coupling as well as the Heisenberg interaction. We make use of these results in order to establish a process for PST between any pair of vertices in hypercube of any dimension.

\section{Memory-enhanced perfect state transfer on hypercubes}\label{sec:hcube}
	


A hypercube of dimension $n$, denoted by $\mathcal{Q}_n$ is a graph on $2^n$ vertices, $n\geq 0$ which can be defined as an $n$-times Cartesian product of the complete graph on $2$ vertices, which is an edge.  Let the vertex set of the complete graph on two vertices be $\{0, 1\}.$ Then the vertices of $\Q_n$ are labeled as the $n$-tuples of $0$ and $1$, that is, the vertex set of $\Q_n$ is $\mathcal{V}_n = \{0, 1\}^n.$ Two vertices $x, y$ of $\Q_n$ are linked by an edge if the Hamming distance of $x, y\in\{0, 1\}^n$ is one.  Two vertices $x=(x_1, \hdots, x_n)$ and $y=(y_1, \hdots, y_n)$ of $Q_n$ are called antipodal if $x_i\neq y_i,$ $i\in\{1, \hdots, n\}.$ Given two vertices $x$ and $y$, we are interested to determine the unique induced sub-hypercube $\Q_d$ for some $d$ of $\Q_n, n\geq d$ such that the vertices $x, y$ of $\Q_n$ are antipodal in $\Q_d$. The following proposition describes the same. 

\begin{proposition}\label{prop:1}
Let $x=(x_1,\hdots,x_n),$ $y=(y_1,\hdots,y_n) \in \mathcal{V}_n,$ the vertex set of $\Q_n.$ Suppose $d=|\{i : x_i\neq y_i, i=1,\hdots, n\}|.$ Then the unique induced sub-hypercube $\Q_d$ of $\Q_n$ with $x, y$ as antipodal vertices of $\Q_d$ is given by the set of vertices
\begin{equation}
\begin{split}
        \mathcal{V}_d=& \{z=(z_1,\hdots,z_n)\in \mathcal{V}_n : z_i=x_i, \,\, \mbox{if} \, x_i=y_i, \, \mbox{and}\\ & z_i\in\{0, 1\} \, \mbox{otherwise}, i=1,\hdots,n \}.
\end{split}
\end{equation}
\end{proposition}

\noindent{\textbf{Proof.}}
Note that if $d=n$ then $x, y$ are antipodal vertices of  $\Q_n.$ Now let $n-d\neq 0.$ Then there are indices $i_1, \hdots, i_{n-d}$ such that $x_{i_j}= y_{i_j},$ $j\in\{1,\hdots, n-d\}.$ Consider the set of vertices 
\begin{equation}
    \begin{split}
        \mathcal{V}_d =& \{z=(z_1,\hdots,z_n)\in \{0, 1\}^n : z_{i_j}=x_{i_j}=y_{i_j}, \\ & j=1,\hdots,n-d\}.
    \end{split}
\end{equation}
Then $|\mathcal{V}_d|=2^d.$ Consider the subgraph of $\Q_n$ that is induced by the vertex set $\mathcal{V}_d.$ Thus $z,w\in \mathcal{V}_d$ are linked by edge in the induced subgraph if and only if $z, w$ are linked by an edge in $\Q_n,$ that is, Hamming distance between $z$ and $w$  is $1.$ The uniqueness of the sub-hypercube $\Q_d$ of $\Q_n$ follows from the fact that it is an induced subgraph of $\Q_n$. This completes the proof.   $\hfill{\square}$

Now we propose a procedure to utilize the induced $d$-dimensional sub-hypercubes of an $n$-dimensional hypercube $\Q_n$, $n\geq d$ for perfect transfer of qubits between any pair of nodes in $\Q_n.$ 

In the following we describe how a classical $n$-bit or $n$-qubit quantum memory address register corresponding to every vertex of $\Q_n$ can help to obtain the desired state transfer between a pair of arbitrary vertices of $\Q_n.$ 
As discussed above, each vertex of $\Q_n$ can be labeled as an element of $\{0, 1\}^n.$ Let $x=(x_1, \hdots, x_n)\in\mathcal{V}_n =\{0, 1\}^n.$ Then consider a classical $n$-bit memory which stores the labelling of a vertex or quantum memory register which store the corresponding $n$-qubit product state $\ket{x}$  for the vertex $x.$ For the classical memory design and architecture, it is a standard procedure that the digital information of the classical bit $0$ or $1$ can be
stored by distinguishing between different values of some continuous physical quantity, such as voltage or current. Besides, many organizations are possible for a memory design  based on the choices of the number of cells in a memory and each cell contains a fixed number of bits, see \cite{tanenbaum2016structured,patterson2016computer}. Setting $n$ bits of data in a cell, a memory of $n2^n$ bits of data is possible having $2^n$ memory addresses. In the present standard of architecture a large number of bits can be stored in a memory, for example, in semiconductor memory devices millions of bits can be stored and accessed \cite{tanenbaum2016structured}. Each cell is labelled with a memory address and a desired collection of addresses can be accessed by a program efficiently \cite{gabbrielli2010programming}. Hence the vertices with desired bits at the desired positions in the cells corresponding to all the vertices of the hypercube can be determined and the vertices of the sub-hypercube $\Q_d$ can be identified.  On the other hand, observe that $\{\ket{x} : x\in\mathcal{V}_n\}$ is the canonical orthonormal basis of $(\C^2)^{\otimes n}.$ Alike a classical memory register, qubits are assumed to be placed in a quantum memory register such that each qubit in a cell can be accessed for a task such as measurement \cite{williams2010explorations}.  Upon measurement of the $d$ qubits in the $n$-qubit register at desired (in which the measurement result matches for both $\ket{x}, \ket{y}$) positions corresponding to every vertex can be obtained. These vertices constitute the vertex set of the desired sub-hypercube $\Q_d.$ However, proposal for the design and architecture of quantum memories is at nascent stage of research, see \cite{brennen2015focus,heshami2016quantum} and the references there.   

Given a pair of vertices $x, y$ of $\Q_n$ finding the vertices in $\Q_n$ with the number of places where the bits of labeling of $x$ and $y$ coincide  may  apparently be seen as a daunting task. However, the largest semiconductor memory chips available in today's technology can hold a few gibibits of data where a gibibit is equivalent to $2^{30}$ bits of data \cite{dawoud2010digital}. Besides, checking the bits at some fixed positions in all the memory registers can be done in parallel in $\mathcal{O}(1)$ time. For a quantum memory register that consists of all the canonical $n$-qubit product states each corresponding to a vertex of the hypercube $\Q_n$, the bottleneck becomes the performance of quantum measurement for each qubit at desired positions of all the $n$-qubit product states.  Finally, we mention that the identification of the desired vertices of the sub-hypercube is independent of the switching procedure described below.

\begin{figure}[hbtp]
    \centering
    \includegraphics[scale=0.43]{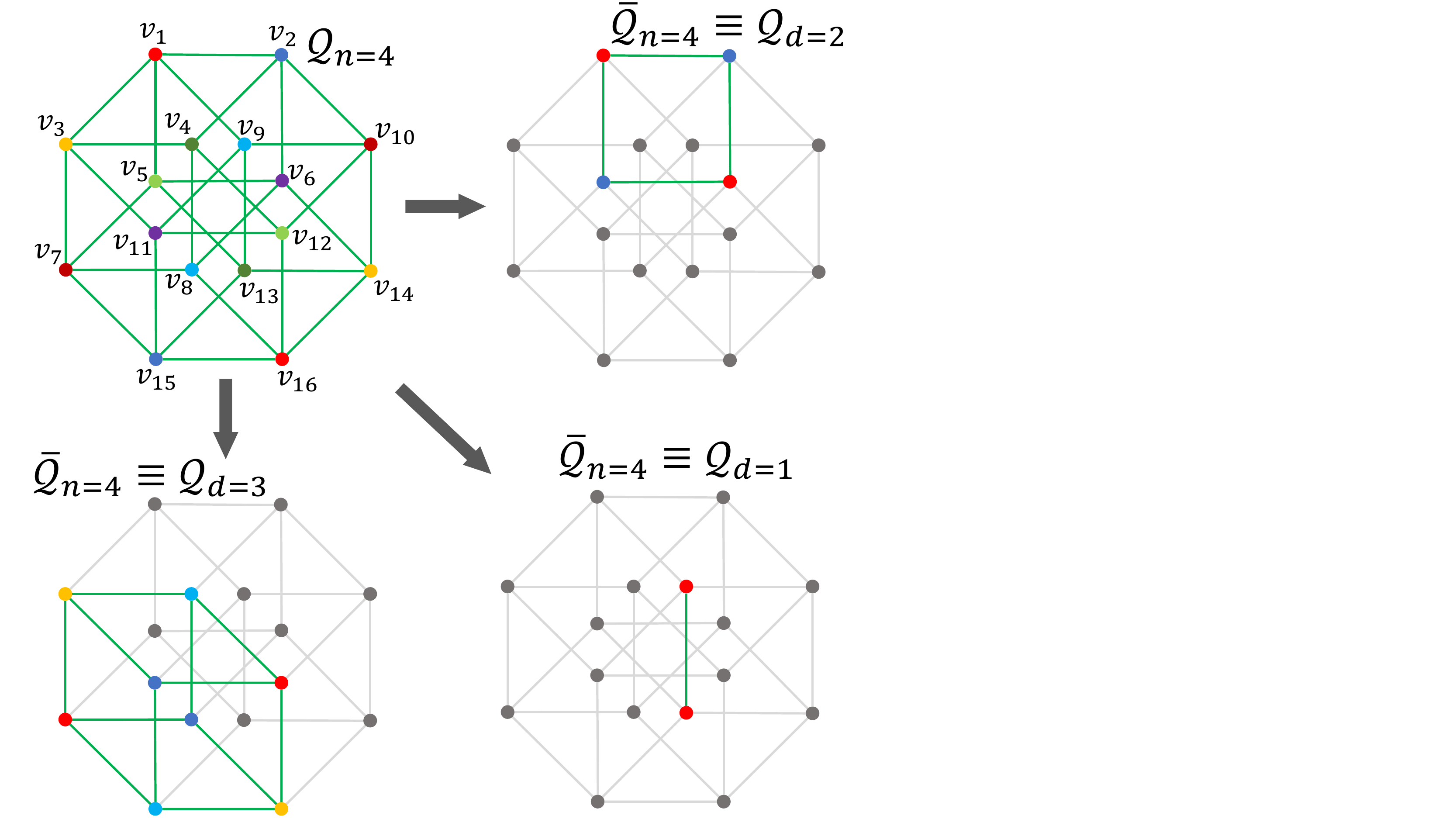}
    \caption{Visual representation of the switching process. Green represents switched-on edges and gray indicates switched-off edges. Isolated vertices are in dark-gray. Any hypercube $\mathcal{Q}_n$ (here $n=4$), under switching can realise various embedded sub-hypercubes ($\overline{\mathcal{Q}}_n$ effectively identical to $\mathcal{Q}_d$ for PST dynamics). Antipodal vertices of the active hypercubes $\Q_d$, are denoted by same the color.}
    \label{fig:hypercubes}
\end{figure}	
Once the list of such vertices is identified, a switching technique is proposed to put in place to create an induced sub-hypercube $\Q_d$ of $\Q_n$ such that $x, y$ are antipodal vertices of $\Q_d$ as defined in Proposition \ref{prop:1},  The switching technique involves tuning of the coupling strength of all the edges in $\Q_n$ that do not belong to the induced sub-hypercube $\Q_d.$ Indeed, once the vertex set $\mathcal{V}_d$ is determined, deactivate all the couplings that incident to any vertex in $\mathcal{V}_n\setminus\mathcal{V}_d.$ We call this process as a \textit{switching process} as it can be interpreted as switching off some edges of $\Q_n$ for communication and once the state is transferred to a desired site, the inactive edges are again switched on for the next job over the hypercube $\Q_n.$ Note that the network state for perfect state transfer is not modified due to the switching technique but it aids to limit the communication between the desired end point vertices of the induced sub-hypercube $\Q_d,$ which can be done by following the procedure proposed in \cite{ref:28}. A proposal for possible physical implementation of the switching technique described above is explained in the next section. Below, we show that  perfect state transfer with unit fidelity is possible in an ideal implementation of the proposed switching procedure.  A clear illustration of the switching process is represented in Fig. \ref{fig:hypercubes}. Starting with the hypercube $\Q_{n=4}$, we label the vertices $v_i \in \mathcal{V}_n$. The vertices are labeled as $v_i=(v_{i_1},v_{i_2},v_{i_3},v_{i_4})$, with $v_1 = (0, \, 0, \,0, \, 0), v_2 = (0, \, 0, \,0, \, 1), v_3 = (0, \, 0, \,1, \, 0), v_4 = (0, \, 0, \,1, \, 1), v_5 = (0, \, 1, \,0, \, 0), v_6 = (0, \, 1, \,0, \, 1), v_7 = (0, \, 1, \,1, \, 0), v_8 = (0, \, 1, \,1, \, 1)$ and $v_{2^4-i+1}=\overline{v}_i=v_i\oplus (1,\, 1,\, 1, \,1),$ the antipodal of $v_i,$ $i=1, \hdots, 8$. When switching to a hypercube, for instance, $\Q_{d=2}$ we have the vertex set $\{ v_1,v_2,v_5,v_6 \}$ in accordance with proposition \ref{prop:1}, with antipodal vertex pair $(v_1,v_6)$ and $(v_2,v_5$). Similarly, it follows for $d=3$ and $d=1$.

Let us denote the graph obtained after applying the switching techniques as $\overline{\Q}_n = \Q_d \sqcup \{v_{i_1}\} \sqcup \hdots \sqcup \{v_{i_{2^n-2^d}}\}$ where $\Q_d$ is a $d$-dimensional induced sub-hypercube of $\Q_n$ and $v_{i_j}, j=1, \hdots, n-d$ are isolated vertices of $\Q_n.$ Let $A=[a_{ij}]$ denote the adjacency matrix associated with $\Q_n.$ Then the adjacency matrix $\overline{A}=[\overline{a}_{ij}]$ corresponding to $\overline{\Q}_n$ is given by $\overline{a}_{ij}=a_{ij}$ if $v_i, v_j\in \mathcal{V}_d,$ and $0$ otherwise.

Then we show that perfect state transfer is possible between any two antipodal vertices of the induced sub-hypercube $\Q_d$ of $\overline{\Q}_n.$ For this, we show that the PST dynamics of $\overline{\Q}_n$ is identical to that of $\Q_d$.

			The graph $\Q_d$ alone in isolation will have perfect state transfer between two vertices $x$ and $y$ at time $t_0$. Also, let $\overline{\Q}_d$ be the fully disconnected graph with with the vertex set $\mathcal{V}_n\setminus\mathcal{V}_d$. Clearly, $\overline{\Q}_d$ has no edge in itself or with $\Q_d$, hence no matrix element via adjacency matrix or Laplacian exists that can exchange an excitation involving these vertices. The PST in isolated $\Q_d$ indicates $|\braket{y|\exp(-i t_0 A(\Q_d)) | x}| = 1$, where $\ket{x}$ and $\ket{y}$ are the canonical basis vectors of $\mathbb{C}^{2^d}$ corresponding to the vertices $x$ and $y$, respectively. 
			
			Since $\overline{\Q}_d$ contains $2^{n}-2^d$ vertices, we know that the graph $\Q_d \sqcup \overline{\Q}_d \equiv \overline{\Q}_n$ has $2^n$ vertices. Corresponding to the vertices $x$ and $y$ define state vectors in $\mathbb{C}^{2^n}$ as $\ket{x'} = \bmatrix{ \ket{x} \\ \mathbf{0}_{|2^n-2^d| \times 1}}$ and $\ket{y'} = \bmatrix{ \ket{y} \\ \mathbf{0}_{|2^n-2^d| \times 1}}$, respectively. Now, the adjacency matrix of $\overline{\Q}_n$ takes the block form
			\begin{equation}
			\label{eqn:PSTfine}
					 A(\overline{\Q}_n) = \bmatrix{ A(\Q_d) & \mathbf{0}_{|2^d| \times |2^n-2^d|} \\ \mathbf{0}_{|2^n-2^d| \times |2^d|} & A(\overline{\Q}_d)}.
					\end{equation}
					However, the adjacency matrix $A(\overline{\Q}_d)$ is a zero matrix because there is no interaction amongst the isolated vertices in $\overline{\Q}_d$. Then the evolution matrix is
					\begin{equation}
					    \exp(-i t_0 A(\overline{\Q}_n)) = \bmatrix{ \exp(-i t_0 A(\Q_d)) & \mathbf{0}_{|2^d| \times |2^n-2^d|} \\ \mathbf{0}_{|2^n-2^d| \times |2^d|} & I }. \\
			\end{equation} 
			Now, the condition for PST between  $x$ and $y$ in $\overline{\Q}_n$ is $	|\braket{y' | \exp(-i t_0 A(\overline{\Q}_n)) | x'}|$
			\begin{equation}
				\begin{split}
			 & = \Bigg|\bra{y}  \bmatrix{ \exp(-i t_0 A(\Q_d)) & \mathbf{0}_{|2^d| \times |2^n-2^d|} \\ \mathbf{0}_{|2^n-2^d| \times |2^d|} & I } \bmatrix{ \ket{x} \\ \mathbf{0}_{|2^n-2^d| \times 1}} \Bigg|\\
				& =\Bigg| \bmatrix{ \bra{y} &  \mathbf{0}_{1 \times |2^n-2^d|}} \bmatrix{ \exp(-i t_0 A(\Q_d)) \ket{x} \\ \mathbf{0}_{|2^n-2^d| \times 1} } \Bigg| \\
				& =| \braket{y | \exp(-i t_0 A(\Q_d)) | x}| = 1.
				\end{split}
			\end{equation} 
		Therefore, the graph $\overline{\Q}_n$ has identical perfect state transfer dynamics as that of the graph $\Q_d$. This guarantees PST in induced sub-hypercube $\Q_d$ under switching. A similar argument can be given considering the Laplacian matrix associated with the hypercube in place of the adjacency matrix which corresponds to Heisenberg model.


	
	
	
\section{Proposal for implementation of switching operation with superconducting qubits}
 \label{sec:implementation}
 

In this section we describe the physical realization for the hypercube switching PST under the XY coupling. The same architecture holds true for the Cartesian products of $P_3$ graph, the path graph on $3$ vertices. 
We propose the implementation of our PST scheme as pretty good state transfer with a fidelity of $\mathcal{F}=1-\mathcal{O}((g/\Delta)^3)$, where $g$ is related to the strength of the qubit-qubit coupling and $\Delta$ is the detuning of qubits. 
Our task is to show support for switching of the hypercube $\Q_n$ to $\overline{\Q}_n$ as desired for any pair of chosen vertices of $\Q_n$ for the task of PST. 
Physical key requirements of our architecture are the following: (a) $n$ nearest-neighbour (NN) interactions to realize any general $\Q_n$ hypercube, (b) Since distant qubits are connected, implementation is not possible in planar integration, $\Q_4$ on wards. Three-dimensional (3D) integration is needed \cite{ref:36,ref:63}, (c) Tunable (switchable) edges as couplings for each pair of nodes (qubits), and (d) High fidelity control over the processor \cite{ref:37}.

\begin{figure}[hbtp]
    \centering
    \includegraphics[scale=0.315]{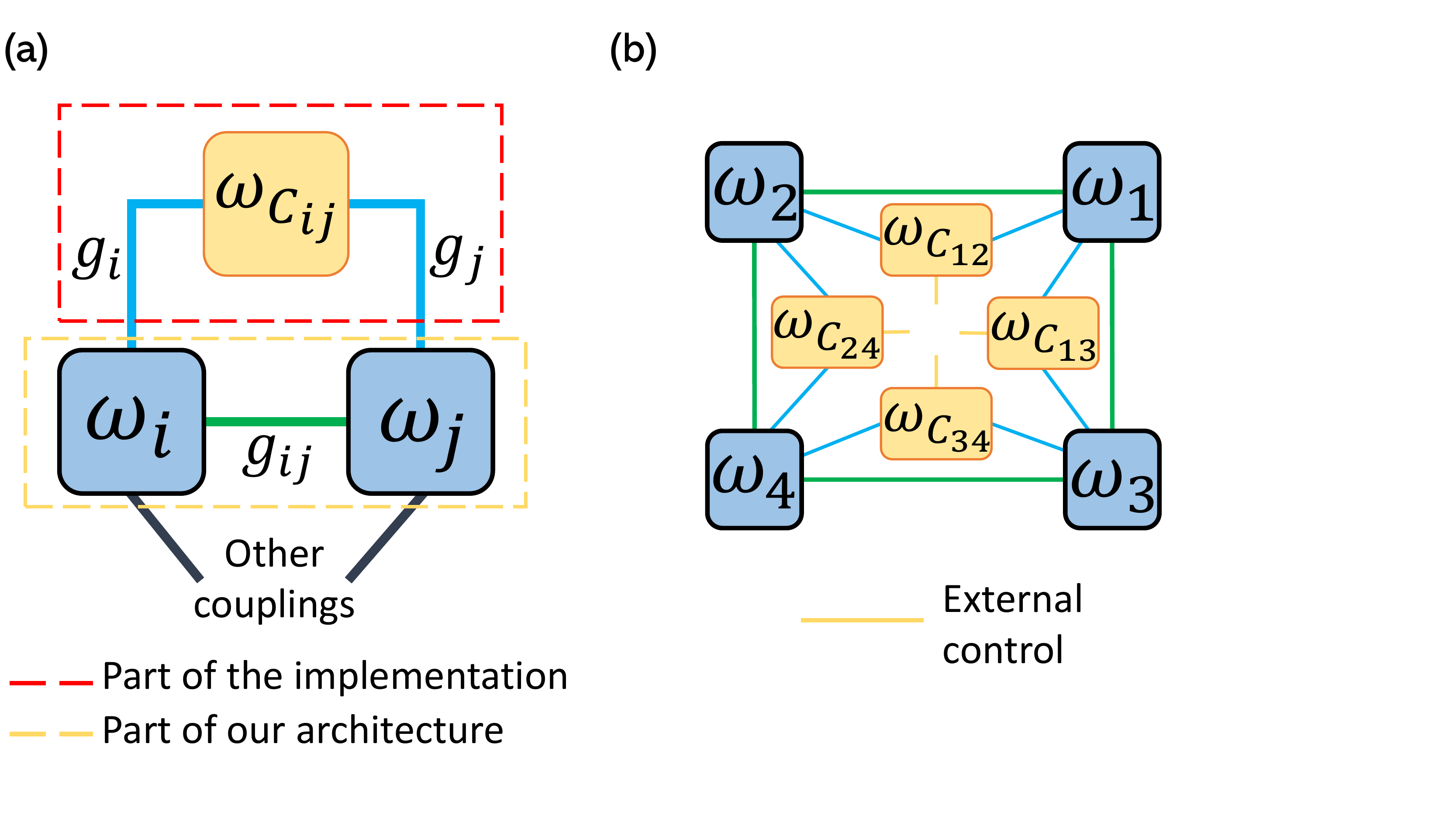}
    \caption{(a) Couplings involved between a pair of interacting qubits, forming an edge in the hypercube $\Q_n$. (b) Network of four qubits forming $\Q_2$. Each ancillary coupler is associated with every edge which is controlled in the experiment.}
    \label{fig:implementation}
\end{figure}

Most of these requirements are hard to realize with the conventional spin lattice where switching and tuning of edge coupling becomes highly challenging.  Moreover, such coupling is a function of the distance between two nodes which is not changeable in practice. 
Instead, to implement our network  we consider a new adjustable coupler that involves ancillary qubits actings as (tunable) couplers \cite{Chen2014, ref:14, ref:31}. See Fig. \ref{fig:implementation}.
 These additional couplers linked to architectural qubits often add complexity, and create a new means for decoherence \cite{Reuther2010}. However, superconducting qubits have already sufficiently long  decoherence times \cite{ref:14, ref:31}, and high fidelity gates are realised using tunable couplers \cite{sung2020realization}.



We consider a general system that consists of $2^n$  qubits for $\Q_n$ with exchange coupling between nearest qubits (which have an edge between them).
In addition, as anticipated,  for having a switchable coupling an extra qubit between them is also needed.
The total number of  these ancillary couplers equal to the number of edges in the network ($n2^{n-1}$ for $\Q_n$).
We use the notation $i$ and $j$ for qubits and $C_{ij}$ for the coupler that connects to these qubits. Both the qubits $i$ and $j$ (with respective Zeeman splittings $\omega_i$ and $\omega_j)$ couple to the auxiliary one $(\omega_{C_{ij}})$ with a strength  $g_i$ $(i = 1, 2,...,2^n)$. In addition they are mutually coupled with transverse coupling strength $g_{ij}$. 
The total Hamiltonian including both physical and ancillary qubits  is  given by:
\begin{equation}
    H_{\Q_n}=\sum_{\langle i, j \rangle}\left( H_i+H_j+H_{c_{ij}}+H_{ic_{ij}}+H_{jc_{ij}}+H_{ij}  \right)
\end{equation}
where  $\langle i, j \rangle :=(i,j)\in E(\Q_n)$ denotes the pair of adjacent vertices in $\Q_n$. Explicitly,
\begin{equation}
\label{eqn:Hfull}
\begin{split}
    \frac{H_{\Q_n}}{\hbar}=& \frac{1}{2}\sum_{i=1}^{2^n}\omega_i\sigma^z_i+\frac{1}{2}\sum_{\langle i, j \rangle}\omega_{C_{ij}}\sigma^z_{C_{ij}} +\sum_{\langle i, j \rangle}g_i \sigma^x_i\sigma^x_{C_{ij}}\\ & +\sum_{\langle i, j \rangle}g_{ij} \sigma^x_i\sigma^x_j
\end{split}
\end{equation}
If we consider capacitive qubit-qubit coupling, which is advantageous in terms of decoherence times, the coupling strengths can be estimated \cite{ref:14}:
\begin{equation}
\label{eqn:git}
    g_j=\frac{1}{2}\frac{C_{jc_{ij}}}{\sqrt{C_jC_{c_{ij}}}}\sqrt{\omega_j\omega_{C_{ij}}}
\end{equation}
and
\begin{equation}
\label{eqn:gijt}
    g_{ij}=\frac{1}{2}(1+\eta_{ij})\frac{C_{ij}}{\sqrt{C_iC_j}}\sqrt{\omega_i\omega_j}
\end{equation}
where $\eta_{ij}=C_{ic_{ij}}C_{jc_{ij}}/C_{ij}C_{c_{ij}}$. Here, $C_\lambda$ is the transmon qubit capacitance, $C_{ij}$ is the direct qubit-qubit coupling capacitance, $C_{ic_{ij}}$ is the qubit-coupler coupling capacitance and $C_{c_{ij}}$ is the coupler capacitance.  
In the following we consider the so called dispersive regime in wich the qubits are well detuned, i.e.  $g_i\ll |\Delta_i|\quad \forall i$, with  $\Delta_i=\omega_i-\omega_{C_{ij}}<0$ the qubit-ancilla detuning.  Therefore,  we can use a perturbation theory  in  $g_i/\Delta_i$. In order to do so, it turns convenient to use the Schrieffer-Wolff  unitary  transformation $U_{SW}=e^\eta$ \cite{ref:24}. In our case,
\begin{equation}
\label{eqn:SW}
\begin{split}
        U_{SW}= & \exp \Bigg(  \sum_{\langle i,j \rangle} \Big[ \frac{g_i}{\Delta_i}\left(\sigma^+_i\sigma^-_{C_{ij}}-\sigma^-_i\sigma^+_{C_{ij}} \right)\\
        &+\frac{g_i}{\Sigma_i}\left(  \sigma^+_i\sigma^+_{C_{ij}}-\sigma^-_i\sigma^-_{C_{ij}} \right) \Big] \Bigg)
\end{split}
\end{equation}
where the second term takes care for the counter-rotating terms and $\Sigma_i=\omega_i+\omega_{C_{ij}}$. The transformation $U_{SW} H U_{SW}^\dagger$ is done up to second order in  $g_i/\Delta_i$.  Besides, it is assumed that the ancillary qubits always remain in their ground state, which is consistent with the condition $\Delta_i=\omega_i-\omega_{C_{ij}}<0$.  After some algebra we end up with the effective qubit-qubit interaction Hamiltonian [see Appendix \ref{app:A}]:
\begin{equation}
\label{eqn:SCtuned}
    \frac{\Tilde{V}}{\hbar}=\sum_{\langle i, j \rangle} \tilde{J}_{ij}\left( \sigma^+_i\sigma^-_j+\sigma^-_i\sigma^+_j \right)
\end{equation}
where 
the effective \textit{tunable} coupling between any two qubits is given by:

\begin{equation}
    \tilde{J}_{ij}\approx \frac{g_ig_j}{2}\left( \frac{1}{\Delta_i}+\frac{1}{\Delta_j}-\frac{1}{\Sigma_i}-\frac{1}{\Sigma_j} \right)+g_{ij}.
\end{equation}

\begin{figure}[hbtp]
    \centering
    \includegraphics[scale=0.27]{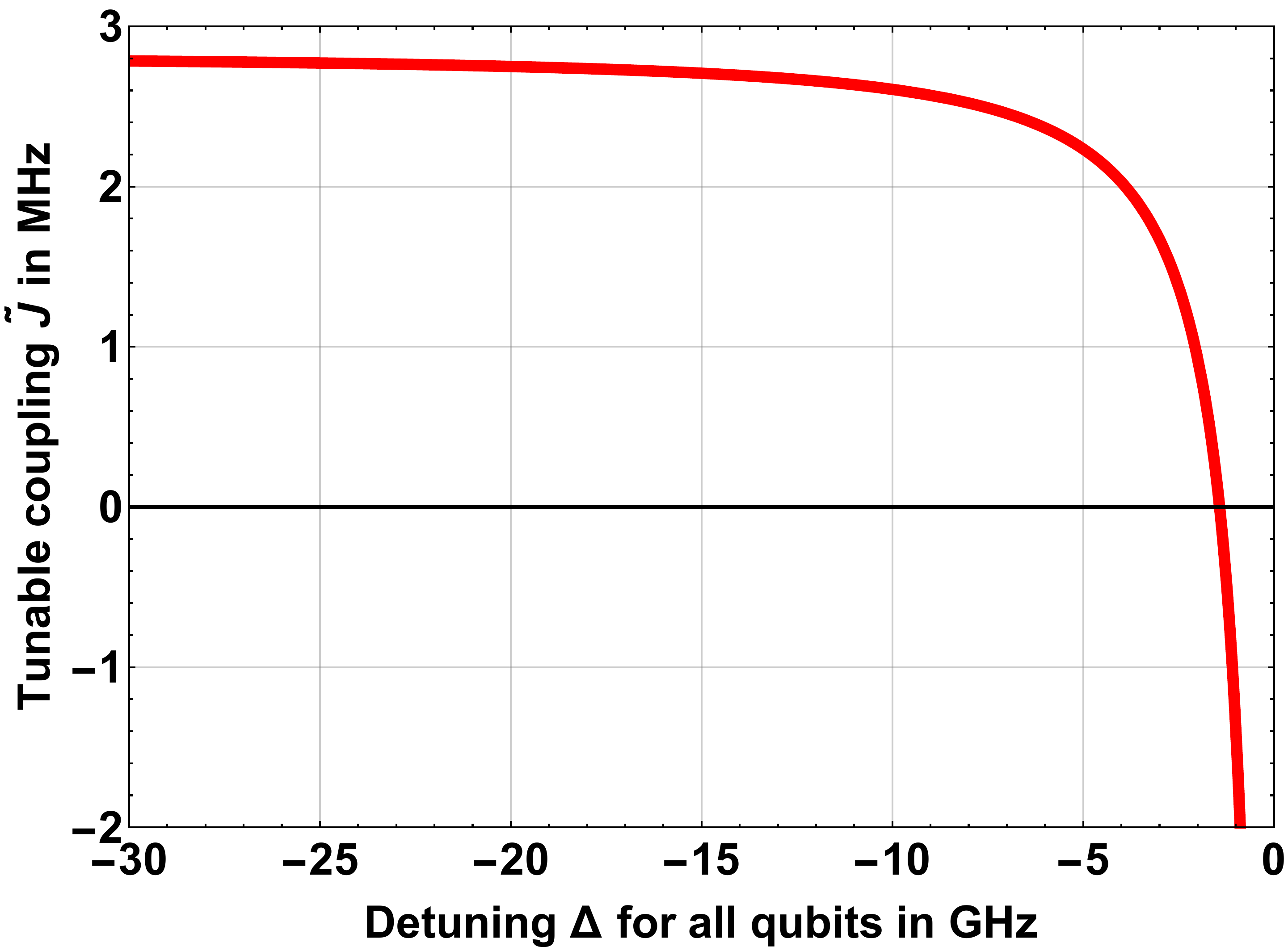}
    \caption{Variation of the dynamic tunable coupling $\tilde{J}$ w.r.t. the detuning $\Delta$ for each qubit. There exists a cutoff value, in this case $\Delta=-1.426$ GHz, corresponding to $\omega^\text{off}_{C_{ij}}=5.426$ GHz. For all configurations, such a cut-off value can always be obtained.}
    \label{fig:coupling}
\end{figure}

Similar effective coupling Hamiltonian based on Cavity and Circuit-QED have been proposed in \cite{ref:22} (scalability has been addressed with experimental concerns using molecular architecture for qubits in superconducting resonators) and \cite{ref:23}. It is clear that the above produces the identical coupling Hamiltonian in Eqn. (\ref{eqn:XYmodel}) which is responsible for qubit-qubit hopping. However, now,  the hopping term  is tunable by setting the desired couplings ($2J_{ij} \longrightarrow \tilde{J}_{ij}$) and detunings. 
In Fig. \ref{fig:coupling} we show how $\tilde{J}_{ij}$ can be altered negative when ancilla coupler frequency is reduced or changed to positive when this frequency is escalated. Therefore, we have some $\omega^\text{off}_{C_{ij}}$ such that $\tilde{J}_{ij}(\omega^\text{off}_{C_{ij}})=0$ within the bandwidth of each coupler. It is shown \cite{ref:14} that this cut-off frequency can be found even in weak dispersive regime with $g_j<|\Delta_j|$. Thus, we obtain the switchable edges with $\omega_{C_{ij}}$ as the parameter. We can simply tune each frequency $\omega_{C_{ij}}$ for each edge $E(i,j)$ to switch it on or off whenever our protocol requires to switch to the graph $\overline{\Q}_n$ from $\Q_n$ and this is essentially a classical operation in experiment. The couplers remain in their ground state throughout the quantum evolution as the effective interaction is only for one quantum exchange between the two qubits which are part of the network for $\Q_n$. 
For the case when all qubits are capacitively coupled  and  $\omega_i=\omega_j=\omega$, using, equation \eqref{eqn:git} and \eqref{eqn:gijt}, the effective coupling is simplified to
\begin{equation}
\label{eqn:coupfinal}
    \tilde{J}_{ij}=\frac{1}{2}\left[ \frac{\omega^2}{ \Delta \Sigma}\eta_{ij}+1 \right]\frac{C_{ij}}{\sqrt{C_iC_j}}\omega.
\end{equation}
In conclusion, to switch on the qubit-qubit coupling as an edge in the network graph, one sets the respective edge coupler's detuning to a reasonable (within the bandwidth) value via $\omega^\text{on}_{C_{ij}}$, yielding a limited $\tilde{J}(\omega^\text{on}_{C_{ij}})$. Then a PST can be performed by modulating only the couplers' frequency to $\tilde{J}(\omega^\text{on}_{C_{ij}})$ for all the edges $E(\Q_d)$. Therefore,
\begin{equation}
   \tilde{V}_{\Q_n} \xrightarrow{\text{switch off } E(\Q_n-\Q_d)}  \tilde{V}_{\overline{\Q}_n\equiv \Q_d}
\end{equation}
and the perfect state transfer time ($t_0$) is related to the dynamical coupling as $t_0=\pi /(2 \tilde{J}^{\text{on}})$.

Fig. \ref{fig:coupling} shows the variation, Eqn. (\ref{eqn:coupfinal}), of the dynamic tunable coupling $\tilde{J}$ with respect to the control parameter $\omega_{C_{ij}}$ through $\Delta$. The reasonable experimental values for the parameters used are \cite{ref:14}: $C_i=70$ fF, $C_j=72$ fF, $C_{c_{ij}}=200$ fF, $C_{ic_{ij}}=4$ fF, $C_{jc_{ij}}=4.2$ fF and $C_{ij}=0.1$ fF. All $\omega_i=\omega_j=\omega=4$ GHz (because all qubits are identical). The fabrication defects and imperfection is accounted in the different values for the capacitance. 
 We have to ensure in the experiment that all detunings are onset to the same value to realise a uniformly coupled qubit hypercube $\Q_d$. If all detunings are not equal this will actually realise a weighted edge hypercube and introduce an error in the fidelity (see Appendix \ref{app:B}). In one way, it can be first compensated by different values of the capacitive couplings involved. For the error that still remains, we can calculate the bound on the fidelity. For the case of example in Fig. \ref{fig:hypercubes} with $\Q_{n=4}$, that the maximum deviation in $\tilde{J}_{ij}$ for each edge is $\pm 0.5 \%$, we have $\mathcal{F}>97.43\%$ (using Eqn. (\ref{eqn:finalerror})).

\section{Conclusion}

We proposed a switching procedure on a memory-enhanced hypercube such that an induced sub-hypercube can be determined with a desired pair of antipodal vertices of the hypercube. Consequently, we showed that the constructed  sub-hypercube enables to provide support for perfect state transfer between any pair of distinct vertices of the hypercube. It is clear that the proposed method can be scaled up to any higher dimensional hypercube except the fact that the access and storage of the labelling of vertices of the hypercube can be done efficiently. A framework of superconducting qubits with tunable couplings is defined for physical implementation of the switching procedure under the XY coupling. It is shown that perfect state transfer between any pair of vertices in a hypercube is possible utilizing the proposed switching procedure. It will be an interesting exposure to extend this method for graphs that have structural support of construction of sub-hypercubes as subgraphs of the graph on any number vertices. We plan to work on this in future. Also, it will be equally interesting to propose a physical implementation of our protocol for the case of Heisenberg interaction where the z-z qubit couplings are required to be tunable. These questions remain to be addressed. 

\begin{appendices}
\numberwithin{equation}{section} 
\section*{Appendix}
\section{ Effective Hamiltonian}
\label{app:A}
In this appendix we derive the effective Hamiltonian (\ref{eqn:SCtuned}) in the main text with the aim of decoupling the ancillary couplers from the computational qubits. Starting with the non-RWA Hamiltonian (\ref{eqn:Hfull}) and employing the Schrieffer-Wolff transformation (\ref{eqn:SW}) we obtain the most general Hamiltonian up to second order in $g_i/\Delta_i$ as

\begin{widetext}
\begin{equation}
\begin{split}
    \frac{\tilde{H}_{\Q_n}}{\hbar}=& \frac{1}{2}\sum_{i=1}^{2^n}\omega_i\sigma^z_i+\frac{1}{2}\sum_{\langle i, j \rangle}\omega_{C_{ij}}\sigma^z_{C_{ij}} +\sum_{\langle i, j \rangle}g_{ij} \sigma^x_i\sigma^x_j+\sum_{\langle i, j \rangle}\Bigg[- \frac{g_i^2}{2}\Bigg(\frac{1}{\Delta_i}-\frac{1}{\Sigma_i}\Bigg)\left( \sigma^+_i\sigma^+_j+\sigma^-_i\sigma^-_j \right)\sigma^z_{C_{ij}}\\
    &+\frac{g_i^2}{2}\Bigg(\frac{1}{\Delta_i}+\frac{1}{\Sigma_i}\Bigg)\sigma_i^z\sigma^+_{C_{ij}}\sigma^+_{C_{ij}}-\frac{g_i^2}{2}\Bigg(\frac{1}{\Delta_i}-\frac{1}{\Sigma_i}\Bigg)\sigma_i^z\sigma^-_{C_{ij}}\sigma^-_{C_{ij}}+\frac{g_i^2}{\Delta_i}\left( \sigma^z_i\sigma^-_{C_{ij}}\sigma^+_{C_{ij}}+\sigma^-_i\sigma^+_i \right)\\
    & +\frac{g_i^2}{\Sigma_i}\left( \sigma^z_i\sigma^-_{C_{ij}}\sigma^+_{C_{ij}}-\sigma^+_i\sigma^-_i\sigma^z_{C_{ij}} \right)\Bigg]+\sum_{\substack{\langle i, j \rangle \\ \langle i, k \rangle}}\Bigg[ \frac{g_i^2}{2}\Bigg(\frac{1}{\Delta_i}-\frac{1}{\Sigma_i}\Bigg)\sigma_i^z\Big( \sigma^-_{C_{ij}}\sigma^+_{C_{ik}}-\sigma^+_{C_{ij}}\sigma^-_{C_{ik}}+\sigma^-_{C_{ij}}\sigma^-_{C_{ik}} \\
    &-\sigma^+_{C_{ij}}\sigma^+_{C_{ik}}\Big)\Bigg]-\sum_{\langle i, j \rangle}\frac{g_ig_j}{2}\Bigg( \frac{1}{\Delta_i}+\frac{1}{\Delta_j}-\frac{1}{\Sigma_i}-\frac{1}{\Sigma_j} \Bigg)\left( \sigma^+_i\sigma^+_j+\sigma^+_i\sigma^-_j+\sigma^-_i\sigma^+_j+\sigma^-_i\sigma^-_j \right)\sigma^z_{C_{ij}}
\end{split}
\end{equation}
\end{widetext}
We now drop the terms which involve double excitation of either the qubits or the couplers and impose the strict condition that all couplers remain in their ground state. This completely decouples the coupler Hamiltonian which can be completely dropped. Further, all the constant energy shift terms can be neglected. This results in the following network Hamiltonian with renormalized frequencies
\begin{equation}
    \begin{split}
        \frac{\tilde{H}_{\Q_n}}{\hbar}=\frac{1}{2}\sum_{i=1}^{2^n}\tilde{\omega}_i\sigma^z_i+\underbrace{ \sum_{\langle i,j\rangle}\tilde{J}_{ij}\left( \sigma^+_i\sigma^-_j+\sigma^-_i\sigma^+_j \right)}_{\tilde{V}}
    \end{split}
\end{equation}
where $\tilde{V}$ is the effective qubit-qubit interaction and
\begin{equation}
    \tilde{\omega}_i\approx \omega_i+g^2_i\Bigg(\frac{1}{\Delta_i}+\frac{1}{\Sigma_i}\Bigg)
\end{equation}
is the new corrected frequency due to the Lamb-shift frequency revealed by the SW transformation and 
\begin{equation}
    \tilde{J}_{ij}\approx \frac{g_ig_j}{2}\left( \frac{1}{\Delta_i}+\frac{1}{\Delta_j}-\frac{1}{\Sigma_i}-\frac{1}{\Sigma_j} \right)+g_{ij}.
\end{equation}
is the effective and dynamic qubit-qubit coupling of the network qubits.

\section{Error calculation for non-ideal switching}
\label{app:B}
Starting from the hypercube $\Q_n$, and switching to a graph $\overline{\Q}_n$ for a desired $\Q_d$, the switching may not be ideal. Meaning that the detuning parameters are not exact. This means that there can be edge strengths which are not exactly zero but coupled with some finite effective strength $\tilde{J}_{ij} \neq 0$ between the qubits $i$ and $j$ that are supposed to be decoupled. And also, there can be edges which are not all identically weighted as $\tilde{J}^{\text{on}}$. Therefore, the effective network graph adjacency matrix is weighted ($A'(\overline{\Q}_n)$). It is important to calculate the error bounds due to these experimental errors over the fidelity of PST. Here we look at such error independently. Eqn. (\ref{eqn:PSTfine}) then takes the form
\begin{equation}
\begin{split}
	A'(\overline{\Q}_n) =& \bmatrix{ A(\Q_d)+E(\Q_d) & E \\ E^T & E(\overline{\Q}_d) } \\ 
	=& \bmatrix{ A(\Q_d) & 0 \\ 0 & 0 } + \bmatrix{ E(\Q_d) & E \\ E^T & E(\overline{\Q}_d)}= \mathcal{U}+\mathcal{E}
\end{split}
\end{equation}
where the $\mathcal{E}=[e_{ij}]$ (with $|e_{ij}|\ll |a_{ij}|$) is a matrix with small norm that captures the effect of unwanted edges. Then from \cite{konstantinov1996improved} we have 
\begin{equation}
	\exp(-it(\mathcal{U}+\mathcal{E}))=\exp(-it\mathcal{U}) + K(-it,\mathcal{E})
\end{equation} where
\begin{equation}
\begin{split}
    K(-it,\mathcal{E}) &= \int_0^t \exp \left(-(it+s)\mathcal{U}\right) \mathcal{E} \exp \left(s(\mathcal{U}+\mathcal{E})\right) ds \\
	&= \sum_{m=1}^\infty K_m(-it,\mathcal{E})
\end{split}
\end{equation}
and
\begin{equation}
\begin{split}
    K_m(-it,\mathcal{E}) &= \int_0^t \exp(-(it+s)\mathcal{U}) G_m(s,\mathcal{E})ds,\\		G_m(s,\mathcal{E}) &= \sum_{r=m-1}^\infty \frac{s^r}{r!} \sum_{i_1+i_2+\hdots+i_m=r-m+1}\Pi_{k=1}^m (\mathcal{EU})^{i_k}
\end{split}
\end{equation} 
Indeed, 
\begin{equation}
    \|K_m(-it,\mathcal{E})\|_2= \mathcal{O}(\|\mathcal{E}\|_2^m)\leq \mathcal{O}(\|\mathcal{E}\|_F^m)
\end{equation}
where $\|M\|_2=\sum_{x\neq 0} \frac{\|Mx\|_2}{\|x\|_2}, \|M\|_F =\left(\sum_{i,j=1}^n |m_{ij}|^2\right)^{1/2}$ for any matrix $M=[m_{ij}]$ of order $n\times n.$ 

Therefore, 
\begin{equation}
\label{eqn:errorin}
 \|K(-it,\mathcal{E})\|_2\leq \sum_{m=1}^\infty \mathcal{O}(\|\mathcal{E}\|_F^m)
\end{equation}
Then modulus inequality reveals that
\begin{equation}
    |\langle y'|e^{-it_0\mathcal{U}}|x'\rangle + \langle y'|K(-it_0,\mathcal{E})|x'\rangle| \geq 1-|\langle y'|K(-it_0,\mathcal{E})|x'\rangle|.
\end{equation}
Using Eqn. (\ref{eqn:errorin}) equipped with Cauchy-Schwarz inequality gives the bound on the effective fidelity as
\begin{equation}
\label{eqn:finalerror}
    \mathcal{F}\geq 1-\sum_{m=1}^\infty \mathcal{O}(\|\mathcal{E}\|_F^m).
\end{equation}
Therefore, all the inaccurate edge strengths give a compromise on the fidelity which is captured by this inequality, by simply knowing the estimate for strength of various effective coupling deviations in $\overline{\Q}_n$.

\end{appendices}

\bibliographystyle{unsrt}
\bibliography{Bibliography.bib}

\end{document}